\documentclass[a4paper,10pt]{article}

\usepackage[english]{babel}
\usepackage[latin1]{inputenc}
\usepackage{graphicx}
\usepackage{epsfig}
\usepackage{hyperref}
\usepackage{wrapfig}

\usepackage[normalem]{ulem}

\newcommand{\pr}{\partial}

\title{Emergence of statistical behavior in many particle mechanical systems: 
Boltzmann's ideas on macroscopic irreversibility}
\author{Navinder Singh\footnote{navinder.phy@gmail.com}\\ Physical Research 
Laboratory, Ahmedabad, India -380009}

\begin{document}

\maketitle

\begin{abstract}
An attempt is made to de-mystify the apparent ``paradox'' between microscopic time revsersibility and macroscopic time  irreversibility. 
It is our common experience that a hot cup of coffee cools down to room temperature and it never 
automatically becomes hot (unless we put that in a microwave for heating or on 
stove etc) and there are numerous examples. This "one sidedness" of physical processes (like cooling of hot cup) is 
in apparent contradiction with the time reversibility of the dynamical equations 
of motion (classical or quantum). The process of automatic heating of a cold cup 
etc is perfectly possible from the dynamical equations perspective. Ludwig 
Boltzmann explained this "one sidedness" of physical processes starting from 
dynamical equations (his H-theorem).  A criticism was raised by Boltzmann's 
contemporaries. The origin of this criticism lies in the  very philosophy of 
"mechanism"  that was very prevalent in the 19th century. Everyone wanted to 
understand physical phenomena through Newtonian mechanics (even J. C. Maxwell 
devised a mechanical mechanism using gears to explain the electromagnetic 
field!). The central issue was how can one obtain this "one sidedness" (time 
irreversiblility) if the underlying dynamical laws are time reversible. 

Number of articles exist in literature on the issue. But those are mathematically oriented and a simple presentation from practical point of view is seriously lacking. 
This article is an attempt to de-mystify this ``paradox'' from  simple and practical point of view.
\end{abstract}

\tableofcontents

\section{The ``paradox'': Microscopic reversibility and macroscopic irreversibility}

"Directionality" or time asymmetry of macroscopic phenomena: There are numerous daily life examples which has "directionality" of time. From our sense perceptions these occur from past to future or in the direction of increasing time. As mentioned in the abstract, consider an example of a hot cup of tea placed on table in a room at ambient temperatures (figure 1). With time, tea and cup cools down to room temperature. The cooling requires the transfer of heat from hotter  tea to colder ambient air. As is well known, at microscopic level, the colder air molecules (with lesser kinetic energy) collide continuously with the hotter outer surface of cup thereby taking the energy from the vibrating molecules at the surface of the cup. The result of these collisions is to reduce the kinetic energy of ``tea molecules'' (mostly $H_2O$, $C_nH_{2n}O_n,~3<n<7$ (sugars), milk (colloid of globules), tea (caffeine, catechins, theanine etc)) and to enhance that of air molecules. Also at the surface of tea lot of activity happens in which hotter water molecules in tea evaporate from the surface and  colder ones in air condense. This process of evaporation and condensation leads to the lowering of the kinetic energy of tea molecules with an end result in which average kinetic energy of tea molecules equals that of air molecules and the system acquires thermodynamic equilibrium\cite{fyn}.

From energetic point of view reverse (heating of tea) is perfectly possible in which energy is transfered from air molecules to tea molecules. 
\begin{figure}[h!]
\includegraphics[width= 5cm, height=4cm]{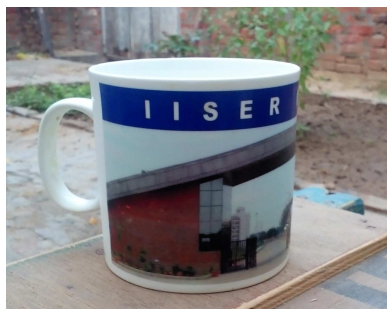}
\caption{Hot tea cup.}
\end{figure}
Thus, from the point of view of first law of thermodynamics, spontaneous heating of a room temperature tea is perfectly possible in which energy is transfered from air to tea. But we never observe this reverse process? 
\begin{figure}
\begin{tabular}{c}
\includegraphics[height = 2.5cm, width = 12cm]{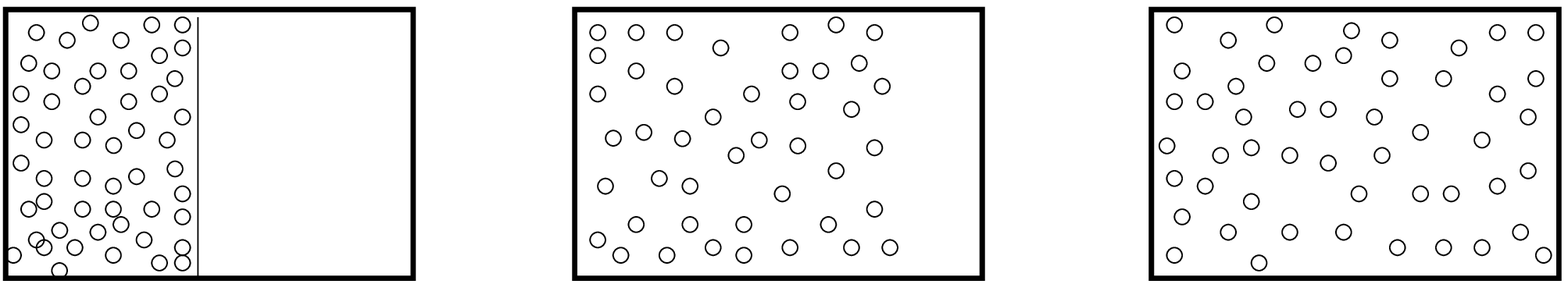}\\
\includegraphics[width= 12cm, height=0.7cm]{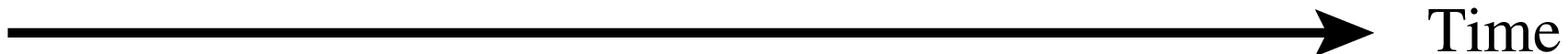}
\end{tabular}
\caption{Gas in one of the compartments of a box}
\end{figure}
One can consider another example. Consider a gas enclosed in one of the compartments of a box (figure 2). When the partition is removed gas expands and fills the whole box. It is never observed that the gas automatically re-occupies the original half at any later time although in the dynamical equations of motion of the gas molecules  there is nothing that prohibits this reverse process. Again, the reverse process is perfectly in accord with the first law of thermodynamics and is in accord with the laws of dynamics. The laws of dynamics or Newton's equations of motion are time symmetric i.e., if at a given instant, velocities of all the molecules are exactly reversed and there is no external influence, the system (gas molecules in a box (figure 2)) will re-trace its microscopic thus ``macroscopic'' trajectory re-filling the first half again.

{\it{The simplest reason why this never happens in practice (although perfectly possible in theory) is that there is ``no superhuman being out there'' that can exactly reverse the velocity vector of all the molecules at a same instant of time. Even if, say, some ``superhuman'' exists and does this, the reverse trajectories of molecules will not be perfect in the sense that very quickly external influences will totally change the course of all the molecules as no system in nature can ever be ideally isolated}}\footnote{Force exerted by the planet Jupiter on a gas molecule (say Nitrogen molecule) in a box/container on earth is roughly $10^{-32}~Newtons$ which sufficient to impart an acceleration of $10^{-6}~m/sec^2$ to the molecule. Under this acceleration path of the molecule can deviate by $\sim 1 micro-meter$ in one second!}. 

Thus, in practice, the question of incompatibility of microscopic time reversibility and macroscopic irreversibility is not relevant. The question becomes relevant when one tries to explain this irreversibility {\it theoretically} that is with a mathematical model. And Boltzmann with a nice mathematical model was able to explain this macroscopic irreversibility with an implicit assumption which, initially, he himself did not recognize.

\section{Resolution of the paradox}

\subsection{Phase space arguments}

One can understand the compatibility of microscopic reversibility with macroscopic irreversibility by first recognizing the important role played by the theory of probability when one considers  relevant observables that are not much dependent  upon the microscopic dynamics\cite{feyn2}. For example, density at a given point in fluid does not depend on how a given molecule is moving about. It is just the average number of molecules in a given volume. {\it Thus observables are highly coarse-grained with respect to microscopic details. Secondly, randomness automatically results when one considers the fact that no system is ideally isolated and motion of any molecule becomes random very quickly due to multitude of forces that it experiences.} One can appreciate ``directionality'' of physical processes by considering the following simple example. 

Let us consider again a box with a molecule or particle in it and mentally visualize a partition in the box. Since the particle is under the action of multitude of random forces, on a time scale much greater than some characteristic time scale of random forces, one has to use probabilistic considerations: 
\begin{enumerate}
\item Probability for the particle to be on a side =$\frac{1}{2}$.
\item Probability for two particles to be on the same side $=\frac{1}{2^2}$
\item Probability for $N$ particles to be on the same side $= \frac{1}{2^N} <<1$ for large $N$.
\end{enumerate}
Thus for large $N$ it is very unlikely that all the particles or molecules spontaneously accumulate in one compartment of the box. The behaviour of systems with few particles is radically different from  those containing very large number (an Avogedro number in ordinary cases) of particles.

The above qualitative considerations can be made quantitative by the following considerations. For this, one has to consider the concept of phase space and microstate. If we have $N$ molecules (point like with no internal degrees-of-freedom) in a box of volume $V$, then there are total $6 N$ degrees-of-freedom. The dynamical state of the whole system can be represented by a point in an extended space of $6 N$ dimensions (called the phase space $\Gamma$). The point in phase space is called the microstate. As the molecules move about under mutual and external interactions the phase point also moves and traverses a trajectory called phase trajectory. Consider now an ensemble of such systems (with the same number of particles in the same volume, and with same ambient temperature and pressure). Now in the phase space there will be a ``swarm'' of phase points (the total number of phase points will be equal to the total number of the members in the ensemble). In this language one recognizes that: It is not that every  microscopic state  at the 
initial time of an ensemble of systems will evolve in accord with experience 
(macroscopic irreversibility), but only a great "majority" of them. The 
"majority" becomes so overwhelming for macroscopic systems that irreversible 
behaviour becomes a certainty. {\it Thus macroscopic irreversibility {\bf emerges} (1) when the number of constituents (atoms/molecules) becomes very large, and (2) the observables are coarse-grained, and (3) for such coarse-grained observables dominant role is not played by the dynamics but by the probabilistic laws.} As explained above the origin of probabilistic laws is in the very fact that no system is ideally isolated and motion of a molecule becomes random due complex external influences.

Boltzmann quantitatively defines these considerations by invoking the concept of entropy. One defines the system at macroscopic level by few parameters i.e., energy (E), volume (V) and the number of particles in the 
system (N). This is called the macrostate $M$ of the system. Clearly a very 
large number of microstates (defined by a point in phase $\Gamma$) corresponds 
to the macrostate $M$ (for example the molecules in the ink droplet can have 
many configurations ("Komplexions" as said by Boltzmann)).  Define $\Gamma_M$ as the region of phase space containing all microstates (for 
specific E, V, and N) called compatible microstates. Define $|\Gamma_M| = 
\int_{\Gamma_M} \prod_{i=1}^N d r_i dp_i,$ as the volume of phase space 
containing all compatible microstates, as a measure of the number of the 
microstates. 

{\it{Above considerations lead Boltzmann to propose
$S_B (X)  = log|\Gamma_M(X)|$ called the Boltzmann entropy, which always increases for irreversible processes, as $|\Gamma_M|$ increases due to above considerations. 
The important point made by him is the identification of this entropy (at 
equilibrium) with the thermodynamic entropy introduced by Clausius, thus 
providing the microscopic foundation of thermodynamics. Thus in an irreversible process both entropies thermodynamic and Boltzmann increase. But there is an essential difference. Boltzmann entropy can be defined for nonequilibrium processes whereas thermodynamic one can only be defined for equilibrium (quasi-static) processes (see for details\cite{gold}). Thus the macroscopic irreversibility is essentially captured by the Boltzmann entropy\cite{leb}.} }

\subsection{Kinetic method}

One can understand irreversibility from  Boltzmann's kinetic method\cite{cer}. This method utilizes the fact that no system is ideally isolated and external forces quickly randomize the molecular motion. The dynamically developed correlations quickly vanish: a fact called molecular chaos (originally ``hypothesis of molecular chaos'' or "Stosszzahl Ansatz"). To understand this consider Boltzmann Kinetic equation for distribution function $f(r,v,t)$ in $\mu-$space ($6-$dimensions) of a molecule:
\[\frac{\pr f(r,v,t)}{\pr t} + v.\frac{\pr f}{\pr r} + a. \frac{\pr f}{\pr v} = 
\int dv_1\int d\Omega g I(g,\theta) [f' f_1' - f f_1].\]
Where $r$ and $v$ are vectors and $f(r,v,t) d^3r d^3v$ gives the fraction of molecules in the $\mu-$ space volume element $d^3r d^3v$ (see for details\cite{cer}). Boltzmann defines the H function as $H(t) = \int dr\int dv f log f$. He proves from his equation that $\frac{d H}{d t} \le 0$, and $H$ is constant when $f' f_1' = f f_1$ and it is given by the Maxwell-Boltzmann distribution.  

In solving his equation he used his famous assumption "Stosszzahl Ansatz" $f_2(r,v_1,r,v_2,t) = f_1(r,v_1,t) f_1(r,v_2,t)$. In this the two particle distribution function is written as the product of single particle distribution functions. The dynamical correlations--developed via collision process which conserve momentum and energy--are contained in the two particle distribution function (and higher order functions). Writing two particle distribution function as a product of single particle distribution functions kills correlations! Although it conserve energy but momentum is randomized. We can now easily see the justification of the assumption made. As mentioned before ``randomness automatically results when one considers the fact that no system is ideally isolated and motion of any molecule becomes random very quickly due to multitude of forces that it experiences''\cite{math}. In view of the great success that Boltzmann equation enjoys the  assumption of "Stosszzahl Ansatz" has posteriori justification too.

\section{Summary}

Thus it is not difficult to see how macroscopic irreversibility emerges even though the fundamental equations that govern the dynamics of constituents (molecules/atoms) obey time reversibility. The simplest reason why this never happens in practice (although perfectly possible in theory) is that there is ``no superhuman being out there'' that can exactly reverse the velocity vector of all the molecules at a same instant of time. Even if, say, some ``superhuman'' exists, and does this, the reverse trajectories of molecules will not be perfect in the sense that very quickly external influences will totally change the course of all the molecules as no system in nature can ever be ideally isolated. Secondly, macroscopic systems involve very large number of constituents, and observables are very coarse with respect of constituents\footnote{For example,  size of a hydrodynamical unit cell (say 1 $mm^3$) is very large as compared to the size of the atom/molecule (sub-nanometers) and time scale on which density fluctuates in that unit cell is much greater than typical molecular time scale (like time between two successive collisions.)}, in such cases probabilistic laws become certainties for all practical purposes. 

When one theoretically (i.e., using a mathematical model) derives irreversibility from microscopic dynamics, one must invoke some ``randomness'' assumption. Boltzmann's "Stosszzahl Ansatz" is a salient example. Randomness automatically results when one considers the fact that no system is ideally isolated and motion of any molecule quickly  becomes random due to multitude of forces that it experiences\cite{math}.

\end{document}